 \documentclass[%
 reprint, apl]{revtex4-1}

\usepackage{booktabs}
\usepackage{xcolor}
\usepackage{graphicx}
\usepackage{graphics}
\usepackage{amsmath}
\usepackage{amssymb}
\usepackage{amsthm,amstext}
\usepackage{amsfonts}
\usepackage{lipsum}
\usepackage{MnSymbol}
\usepackage{mathrsfs}
\usepackage{stmaryrd}
\usepackage{latexsym}
\usepackage[applemac]{inputenc}
\usepackage{accents}

\usepackage[bbgreekl]{mathbbol}
\DeclareSymbolFontAlphabet{\mathbbm}{bbold}
\DeclareSymbolFontAlphabet{\mathbb}{AMSb}
\DeclareMathAlphabet\mathbfcal{OMS}{cmsy}{b}{n}

\newcommand\be{\textbf{\emph{e}}}
\newcommand\bd{\textbf{\emph{d}}}

\newcommand\bE{\textbf{\emph{E}}}
\newcommand\bD{\textbf{\emph{D}}}

\newcommand\F{\textbf{F}}

\newcommand\C{\textbf{C}}
\renewcommand\S{\textbf{S}}

\renewcommand\d\delta
\newcommand\D\Delta
\newcommand\e{\varepsilon}

\newcommand\bsigma{\boldsymbol{\sigma}}

\newcommand\tr{\text{tr}}

\newcommand\beq{\begin{equation}}
\newcommand\beqn{\begin{eqnarray}}
\newcommand\eeq{\end{equation}}
\newcommand\eeqn{\end{eqnarray}}

\newcommand{\trsp}{^{\hspace{-1pt}\textsf{T}\hspace{-1pt}}}
\newcommand{\mtrsp}{^{-\hspace{-0.5pt}\textsf{T}\hspace{-1pt}}}

\usepackage[normalem]{ulem}

\begin{document}

\title{Fine tuning the electro-mechanical response of dielectric elastomers}

\author{Giuseppe Zurlo}%
\email{giuseppe.zurlo@nuigalway.ie}
\affiliation{School of Mathematics, Statistics and Applied Mathematics, NUI Galway, University Road, Galway, Ireland.}%

\author{Michel Destrade}
\email{michel.destrade@nuigalway.ie}
\affiliation{School of Mathematics, Statistics and Applied Mathematics, NUI Galway, University Road, Galway, Ireland.}

\author{Tongqing Lu}
\email{tongqinglu@xjtu.edu.cn}
\affiliation{State Key Lab for Strength and Vibration of Mechanical Structures, Department of Engineering Mechanics, Xi'an Jiaotong University, Xi'an 710049, China}

\date{\today}


\begin{abstract}	

We propose a protocol to model accurately the electromechanical behavior of dielectric elastomer membranes using experimental data of stress-stretch and voltage-stretch tests.
We show how the relationship between electric displacement and electric field can be established in a rational manner from this data.
Our approach demonstrates that the {\it ideal dielectric model}, prescribing linearity in the purely electric constitutive equation, is quite accurate at low-to-moderate values of the electric field and that, in this range, the dielectric permittivity constant of the material can be deduced from stress-stretch and voltage-stretch data. 
Beyond the linearity range, more refined couplings are required, possibly including a non-additive decomposition of the electro-elastic energy. 
We also highlight that the presence of vertical asymptotes in voltage-stretch data, often observed in the experiments just prior to failure, should not be associated with strain stiffening effects, but instead with the rapid development of electrical breakdown. 
\end{abstract}
 
 
\maketitle

A major challenge in the modelling of soft dielectrics membranes with compliant electrodes comes from attempting to deduce by experimental tuning the total energy density function $\Omega(\F,\bE)$, which depends on the deformation gradient $\F$ and on the Lagrangian electric field $\bE$. 
An accurate knowledge of $\Omega$ is essential to formulate correctly the equations of electromechanical equilibrium, giving the Piola-Kirchhoff stress as $\S=\partial_{\F}\Omega$ and the Lagrangian electric displacement vector as $\bD=-\partial_{\bE}\Omega$. 
Then the total Cauchy stress follows as $\bsigma=J^{-1}\S\F\trsp$ with $J=\det\F$, and the Eulerian electric field and displacement vector as $\be=\F\mtrsp\bE$ and $\bd=\F\bD$, respectively.

For isotropic and incompressible dielectrics, the electro-elastic energy depends on $(\F,\bE)$ through five scalar invariants \cite{DoOg05}, for instance: $I_1=\tr \C$, $I_2=\tr\C^{-1}$, $I_4=\bE\mathbf\cdot\bE$, $I_5=\C^{-1}\bE\mathbf\cdot\bE$, $I_6=\C^{-1}\bE\mathbf\cdot\C^{-1}\bE$, where $\C=\F\trsp\F$ is the right Cauchy-Green tensor. 
Noting that $I_5=||\be||^2$, a fairly general and common assumption made in the literature is that the electro-elastic energy is split additively into
\beq\label{energy}
\Omega = W(I_1,I_2) - \psi(\sqrt{I_5})
\eeq
where $W$ is a purely elastic term and $\psi$ is an electro-elastic term. 
This expression encompasses, for example, the so-called {\it ideal dielectric model} \cite{ZhSu07} with $\psi(x)=\e x^2/2$, where $\e$ is the dielectric constant, and the \emph{polarization saturation model} \cite{Li2011} with $\psi(x) = (d_s^2/\e)\ln(\cosh(\e x/d_s))$, where  $d_s$ is the saturation value of the Eulerian electric displacement. 

In this paper we leave $\psi$ unspecified and show that this function can be completely determined from independent sets of stress-stretch and voltage-stretch experimental data. 

Assume that the dielectric membrane is a plate of uniform reference thickness $H$, with compliant electrodes on its upper and lower faces. 
For equi-biaxially deformed states the deformation gradient is $\F=\text{diag}(\lambda,\lambda,\lambda^{-2})$ where $\lambda$ is the in-plane stretch, the Lagrangian electric field is $\bE=(0,0,E)$ and the displacement vector is $\bD=(0,0,D)$, where the non-vanishing components are along the  direction perpendicular to the membrane mid-surface. 
Note that $E=V/H$, where $V$ is the controlled voltage,
and that $e=\lambda^2 E=\sqrt{I_5}$ and $d=\lambda^{-2}D$. 

Specialising the energy \eqref{energy} to equi-biaxial states, we introduce  $\omega(\lambda,E)=w(\lambda)-\psi(\lambda^2 E)$, where $w(\lambda)=W(2\lambda^2+ \lambda^{-4},2\lambda^{-2}+ \lambda^4)$ for the purely elastic part. Electro-mechanical equilibrium in the presence of a pre-stretch $\lambda_p$ reads: $\partial_{\lambda}\omega(\lambda,E) = \partial_{\lambda}\omega(\lambda_p,0)$, or 
\beq\label{eq}
s(\lambda) - \lambda E \psi'(\lambda^2 E) = s(\lambda_p)
\eeq
where $s(\lambda)=w'(\lambda)/2$ is the purely mechanical in-plane stress (and $s(\lambda_p)$ is the mechanical stress required to pre-stretch the plate). Furthermore, the electric displacement reads $D = - \partial_E\omega = \lambda^2 \psi'( E \lambda^2)$, which can be conveniently recast in Eulerian form as
\beq\label{dd}
d=\psi'(e). 
\eeq
Remarkably, Equations \eqref{eq} and \eqref{dd}  show that if the purely mechanical response of the polymer  is known through the function $s(\lambda)$, then the relation between $d$ and $e$ can be determined by an independent set of voltage-stretch data. 
More precisely, once various sets of experimental data $\{\lambda_p,\lambda,E\}$ are recorded, then the resulting curve $d(e)$ can be determined by interpolating the points
\beq\label{main}
d=\frac{s(\lambda) - s(\lambda_p)}{\lambda E},\qquad e=\lambda^2 E. 
\eeq
We emphasise that because $d$ depends only on $e$ according to \eqref{dd}, the purely electrical constitutive equation in Eulerian form must be independent of the specific values of $\lambda$, $\lambda_p$ and $E$. 
If the data does not show this independence, then the  assumption of additivity \eqref{energy} must be abandoned.
\begin{figure}[!th]
\includegraphics[scale=.48]{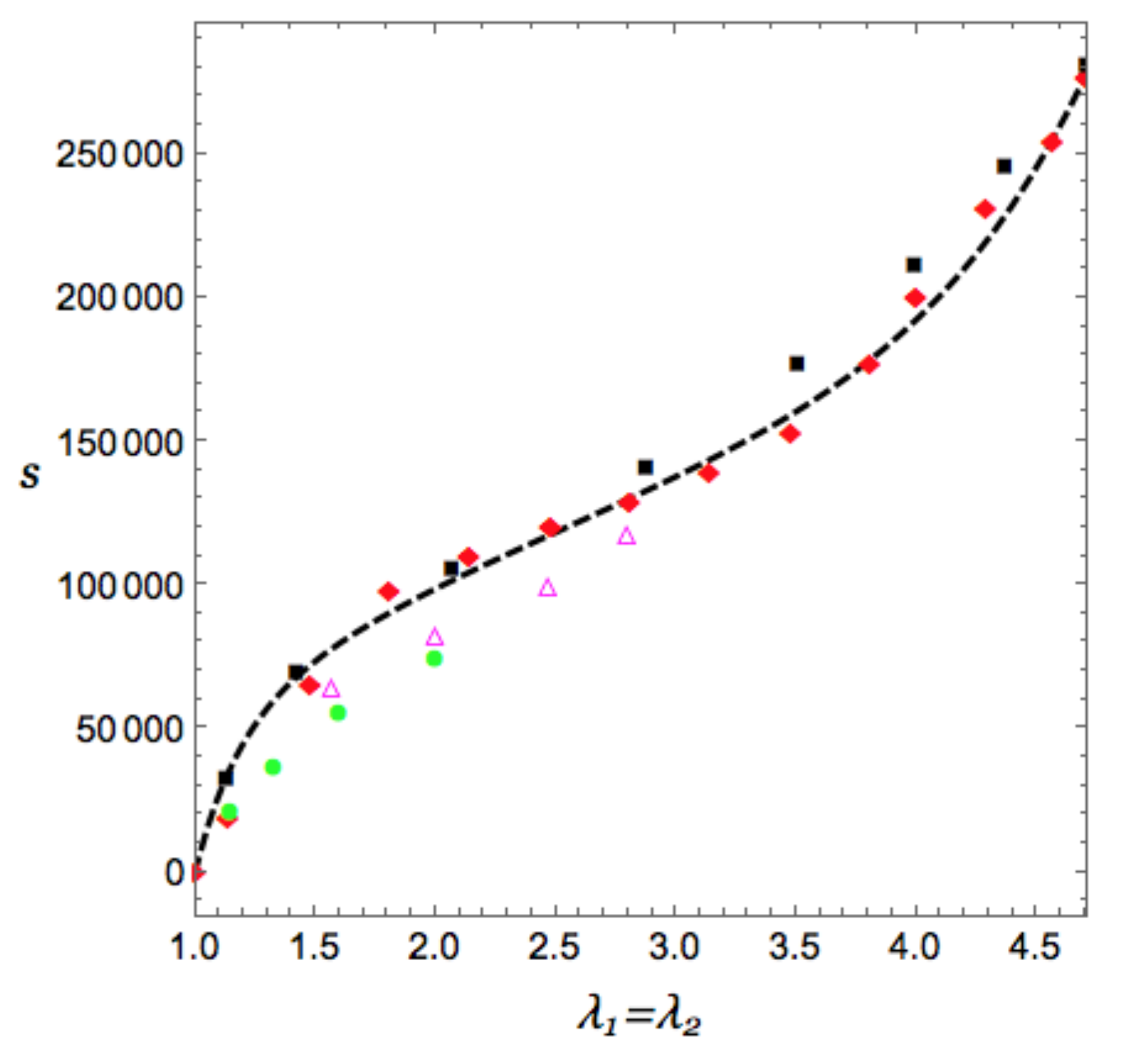}
\caption{\label{mec} Mechanical tuning of the 3M VHB 4905 in equibiaxial tension. 
Black squares taken from \cite{SuoGiant}, red diamonds refer to our own experiment. 
For information, we also give the stress-stretch pairs deduced from the voltage-stretch experiments of \cite{Lu2016} (green circles) and \cite{SuoGiant} (magenta triangles) when the voltage is zero.
Dashed curve: best-fit function $s(\lambda)$.}
\end{figure}

The first step towards the derivation of the purely electric constitutive equation through \eqref{main} is to determine the mechanical response of the polymer.  
Here we focus on the popular dielectric acrylic elastomer VHB4905 (3M Center, St. Paul, MN 55144-1000). This is a viscoelastic material, but here we focus on rate-independent effects.

To provide a robust tuning of the material response, we report two sets of data, see Fig.\ref{mec}. 
The first set comes from direct equi-biaxial tensile tests carried out in our Soft Machine Laboratory (Xi'an Jiaotong University): we used disks with initial radius $R=17.5$mm and thickness $H=0.5$mm, with a loading rate of $2$mm/min. 
The second set of data is extracted from the paper by Huang et al.\cite{SuoGiant} (same $R$ and $H$, loading rate: $11.0$g/90s, which was the slowest rate reported). 

\begin{figure}[!th]
\includegraphics[width=\columnwidth]{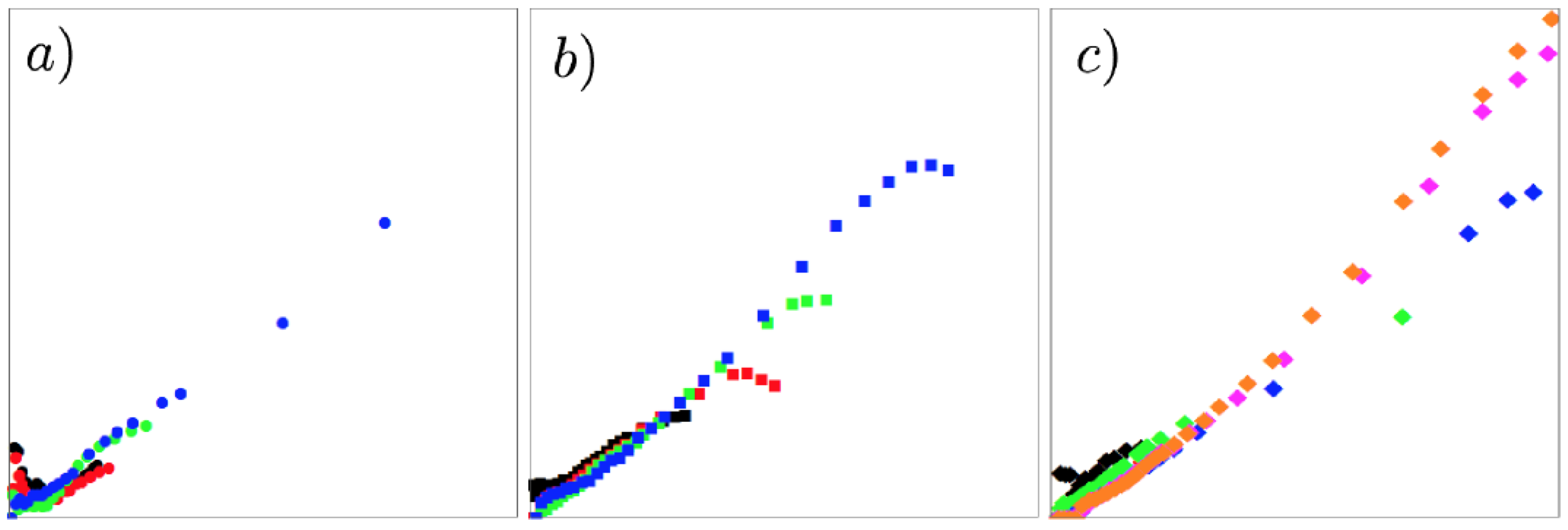}\\
\includegraphics[scale=.54]{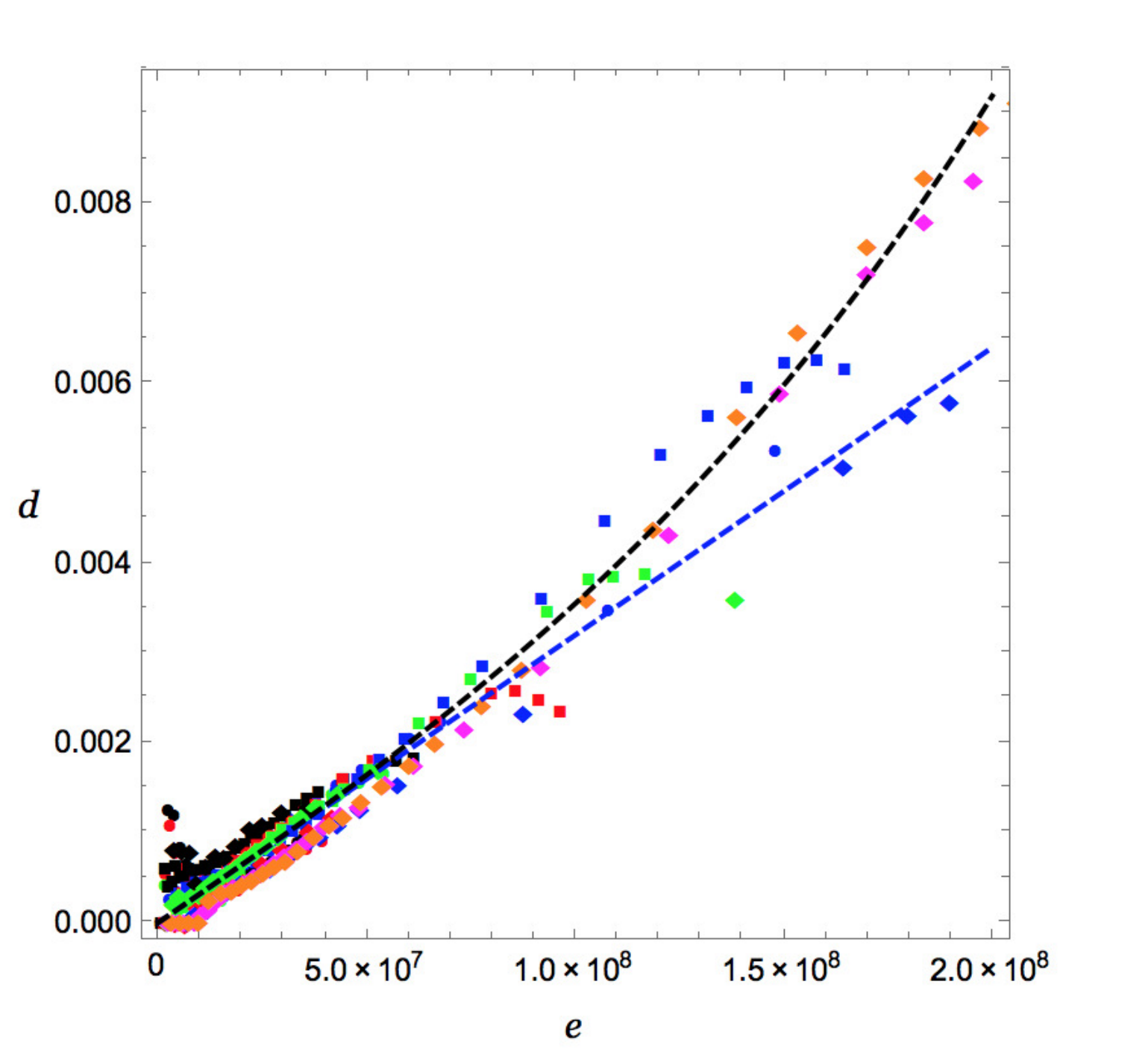}
\caption{\label{de}
Pairs $(d, e)$ in (C/m$^2$, V/m) deduced from Eq.\eqref{main} once the mechanical response $s(\lambda)$ has been determined.
The pairs characterise the purely electrical response of the polymer for three sets of experiments. 
Circle points in $a)$ are obtained from \cite{Lu2016}, where the colors black, red, blue, green correspond to the pre-stretches $\lambda_p=(1.2,1.3,1.6,2)$, respectively. 
Square points in $b)$ are obtained from \cite{SuoGiant}, where the colors black, red, blue, green correspond to the pre-stretches $\lambda_p=(1.6,2,2.5,2.8)$, respectively. 
Diamond points in $c)$ are obtained from \cite{Lu2012}, where the colors black, red, blue, green, magenta, orange correspond to the pre-stretches $\lambda_p=(1.2,1.6,1.8,2,2.5,2.9)$, respectively. 
The points are all aggregated in the main picture below, where the blue and black dashed curves represent linear and cubic interpolations of the data, respectively.}
\end{figure}

We notice that the material breaks before it can exhibit the strain-stiffening effect due to limiting chain extensibility. 
It is thus not appropriate to use a Gent or an Arruda-Boyce model \cite{HoSa02} here, because their stiffening parameter cannot be determined properly. 
Instead, we find that a good fit is obtained with  the following strain energy density (based on the $I_1-$model of Lopez-Pamies \cite{LoPa10}),
\beq\label{smech}
W(I_1,I_2) = \frac{c_1}{2}(I_1^{\alpha}-3^{\alpha}) + \frac{c_2}{2}(I_2^{\beta}-3^{\beta}) 
\eeq
with $c_1=75\,400$Pa, $c_2=0.23$Pa, $\alpha=0.87$ and $\beta=2.2$, see the corresponding graph of $s(\lambda)$ on Fig.\ref{mec}. 
The initial shear modulus for this model is computed as $\mu = 56\, 869$Pa from the formula $\mu := 2(\partial W/\partial I_1 + \partial W/\partial I_2)$ at $I_1=I_2=3$.

We now move on to modelling the electromechanical behavior. 
We consider the $(E, \lambda)$ data from equi-biaxial voltage-stretch experiments conducted on VHB4905 disks with 14 different levels of pre-stretch, taken from the articles \cite{Lu2016,SuoGiant,Lu2012}. 
Fig.\ref{de} shows the distributions of the pairs $(d,e)$ deduced from \eqref{main} (now that $s(\lambda$) is known).

The insets $a),b),c)$ refer to  \cite{Lu2016,SuoGiant,Lu2012}, respectively.
The $(d,e)$ pairs show a late ``electric softening'' behavior, in the sense that the $d-e$ plots  have a decreasing slope just prior to the electrical breakdown. In some cases, $b)$ in particular, breakdown occurs after the recording of a peak; in $a)$ and $c)$ it occurs before a peak can be recorded
(this could be explained by how fast the final expansion occurs and how difficult it is to control it and to measure its progression with increasing dead-loads). 

\begin{figure}[!thb]
\includegraphics[width=\columnwidth]{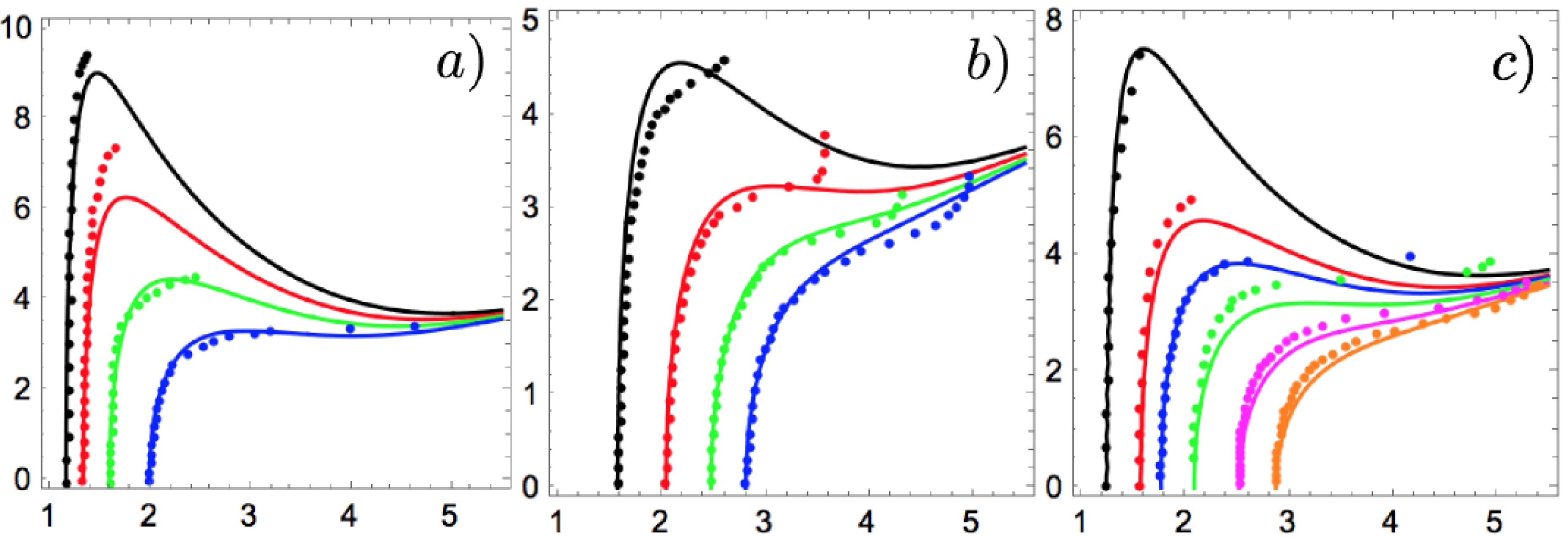}\\
\includegraphics[scale=.45]{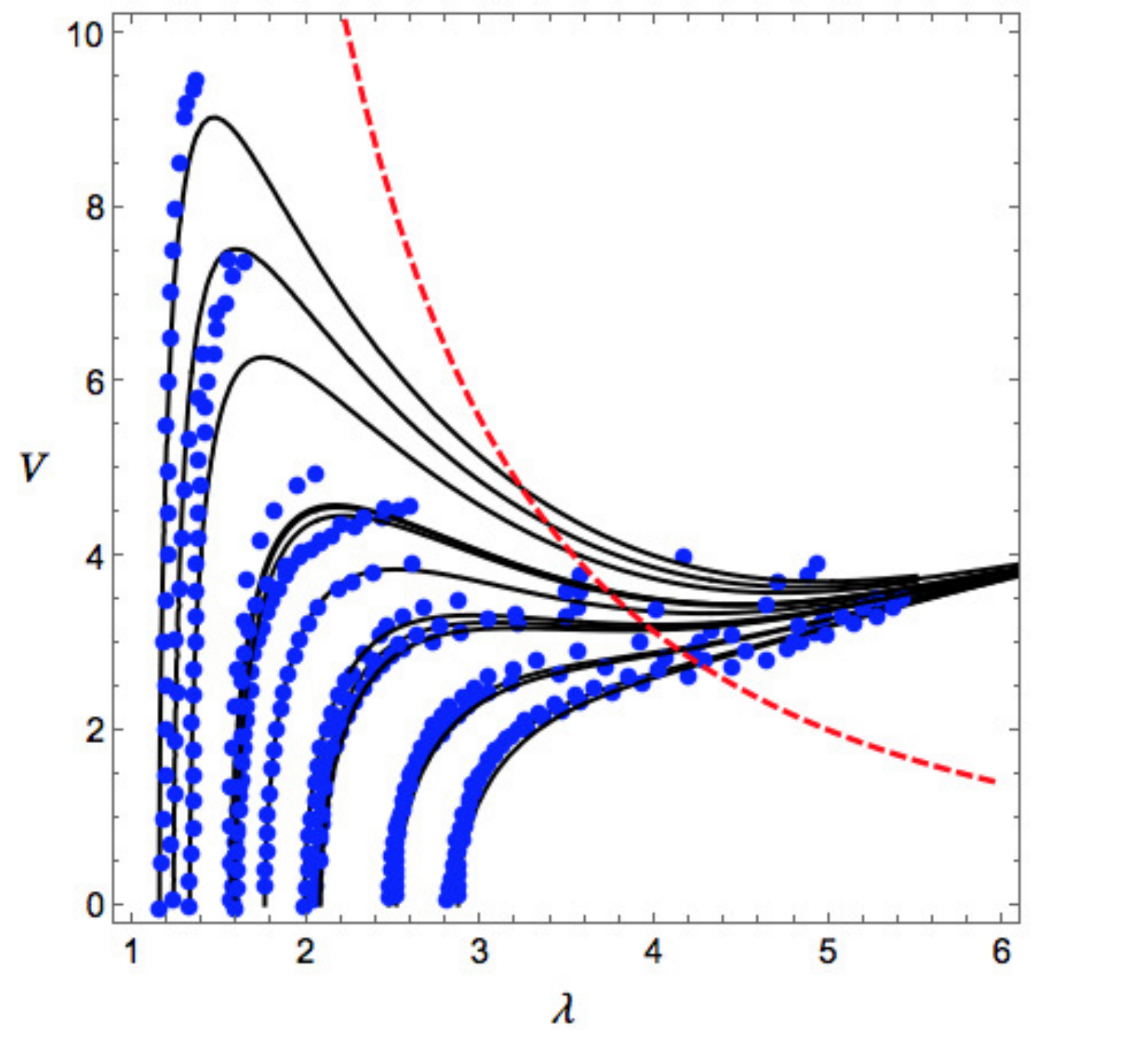}
\caption{\label{all} Compound picture of experimental data (dots) vs theory (solid curves) for all the prestretch levels taken in consideration in this manuscript. The three insets $a),b),c)$ refer to the three sets of experiments \cite{Lu2016},\cite{SuoGiant},\cite{Lu2012}, respectively. Voltage is measured in kV. Solid curves are obtained by using the mechanical response \eqref{smech} together with the cubic electrical response \eqref{dcub}. The red dashed curve, corresponding to $e=10^8$Vm$^{-1}$, represents the limit of applicability of the ideal dielectric model for the VHB4905.}
\end{figure}
The compound picture in Fig.\ref{de} further suggests that the softening branches of the 14 sets of $(d,e)$ pairs may be seen as {\it outliers} with respect to a collective ``safe'' behavior of the polymer. 
This safe branch, which we deem to be located in the $0 \le e \le10^8$Vm$^{-1}$ window, is fitted by the linear interpolation
\begin{equation}
d_{\text{lin}}=\e e, \qquad\e=3.2\times 10^{-11}\,\,\text{F}\text{m}^{-1}
\end{equation}
We conclude that in that range, \emph{VHB4905 behaves as an ideal dielectric}. 
We note that the value of the deduced dielectric constant is of the same order of magnitude, but lower than the usual value $\approx 3.8\times 10^{-11}\,\,\text{F}\text{m}^{-1}$ proposed in literature for this material, see e.g.  \cite{Lu2016,SuoGiant,Lu2012}.

The ideal dielectric paradigm is widely used in the literature on dielectric elastomers, but its direct proof from actuation experiments is quite rare. 
Here we used a rational approach to deduce its validity indirectly. 

Beyond that range, a cubic approximation
\beq\label{dcub}
d_{\text{cub}} = \e e + a e^3
\eeq
with $\e$ as above and $a=3.5\times 10^{-28}$ Fm$^2$V$^{-3}$  fits the data better; 
it gives $\psi = \e I_5/2 + a I_5^2/4$ for the electro-elastic part of the free energy.
Of course, it is debatable whether the cubic is able to capture the scatter observed in Fig.\ref{de} at the higher values of the Eulerian electric field $e$.
As we do not have enough data to conduct a statistical analysis, we leave this question open. 

An alternative point of view is to conclude that the scatter is too large and that  a more sophisticated form of free energy than in Eq.\eqref{energy} must be adopted.
We tried to replace $\psi$ with linear combinations of $I_4$, $I_5$, $I_6$ but they also lead to $d$ being a function of $e$ only, which excludes scatter in the $d-e$ plots. 
Possibly, the additive split of a function of $I_1$, $I_2$ only, and a function of $I_4$, $I_5$, $I_6$ only, has to be abandoned to model the electro-mechanical behavior of VHB4905 over the entire range of stretches and voltages up to breakdown.

Electric softening occurs at high values of the electric field but without any matching behavior (such as strain-softening/stiffening) in the purely mechanical response of the polymer. 
On the other hand, we discard the possibility that |at least for this material| electric softening can be explained in terms of ``polarization saturation'' in the electrical response of the polymer \cite{Li2011}, because none of the $d-e$ curves exhibit saturation. 
Instead we propose that electric softening is simply a global manifestation of a local failure process.
For this reason, this effect should not be regarded (nor modelled) as a constitutive feature of the material.  

We finally show that the voltage-stretch response of the polymer can be described and modelled by using the purely mechanical response function $s(\lambda)$ and the purely electrical response \eqref{dcub}. The resulting theoretical curves show an excellent prediction of the experimental data in a wide range of pre-stretches.  Fig.\ref{all} shows the overall data of the 14 levels of prestretch from the three different articles  \cite{Lu2016,SuoGiant,Lu2012}, together with our theoretical predictions.

\bigskip

{\bf Acknowledgments.} 
The work of T.L. was supported by the NSFC (No. 11772249). G.Z. gratefully acknowledges the hospitality of the PMMH-ESPCI ParisTech and the support of the Italian GNFM (Gruppo Nazionale di Fisica Matematica).
The authors thank the organisers of the 2016 EMI International Conference of ASCE in Metz, France, where preliminary ideas for this paper were discussed.  
They also thank Ruisen Yang for providing  some of the experimental data in Fig.1.


\end{document}